\documentclass[sigconf]{acmart}

\pdfoutput=1
\usepackage{caption}
\usepackage{subcaption}
\usepackage{listings}

\usepackage{enumitem}

\newcommand{\user}{\emph{user}}
\newcommand{\admin}{\emph{admin}}

\definecolor{dkgreen}{rgb}{0,0.6,0}
\definecolor{gray}{rgb}{0.5,0.5,0.5}
\definecolor{mauve}{rgb}{0.58,0,0.82}

\newcommand\YAMLcolonstyle{\color{red}\small\ttfamily}
\newcommand\YAMLkeystyle{\color{black}\small\ttfamily}
\newcommand\YAMLvaluestyle{\color{blue}\small\ttfamily}

\lstdefinelanguage{yaml}
{  
  keywords={true,false,null,y,n},  
  keywordstyle=\color{darkgray}\bfseries,
  basicstyle=\YAMLkeystyle,                                 
  sensitive=false,
  comment=[l]{\#},
  morecomment=[s]{/*}{*/},
  commentstyle=\color{purple}\ttfamily,
  stringstyle=\YAMLvaluestyle\ttfamily,
  moredelim=[l][\color{orange}]{\&},
  moredelim=[l][\color{magenta}]{*},
  moredelim=**[il][\YAMLcolonstyle{:}\YAMLvaluestyle]{:},   
  morestring=[b]',
  morestring=[b]",
  literate =    {---}{{\ProcessThreeDashes}}3
                {>}{{\textcolor{red}\textgreater}}1     
                {|}{{\textcolor{red}\textbar}}1 
                {\ -\ }{{\ttfamily\ -\ }}1,
}

\AtBeginDocument{%
  \providecommand\BibTeX{{%
    \normalfont B\kern-0.5em{\scshape i\kern-0.25em b}\kern-0.8em\TeX}}}

\setcopyright{none}
\renewcommand\footnotetextcopyrightpermission[1]{}
\begin{document}

\title{A Scalable and Cloud-Native Hyperparameter Tuning System}


\author{Johnu George}
\email{johnugeo@cisco.com}
\affiliation{%
  \institution{Cisco Systems,\\
  Bangalore, India}
}

\author{Ce Gao}
\email{gaoce@caicloud.io}
\affiliation{%
  \institution{Caicloud,\\
  Shanghai, China}
}

\author{Richard Liu}
\email{richards.liu@gmail.com}
\affiliation{%
  \institution{Google,\\
  Mountain View, USA}
}

\author{Hou Gang Liu}
\email{liuhgxa@gmail.com}
\affiliation{%
  \institution{IBM,\\
  Xi'an, China}
}

\author{Yuan Tang}
\email{terrytangyuan@gmail.com}
\affiliation[obeypunctuation=true]{%
  \institution{Ant Financial Services Group, Sunnyvale, USA}
}

\author{Ramdoot Pydipaty}
\email{rapydipa@cisco.com}
\affiliation{%
  \institution{Cisco Systems,\\
  Bangalore, India}
}

\author{Amit Kumar Saha}
\email{amisaha@cisco.com}
\affiliation{%
  \institution{Cisco Systems,\\
  Bangalore, India}
}

\renewcommand{\shortauthors}{Johnu et al.}
\begin{abstract}
In this paper, we introduce Katib: a scalable, cloud-native, and production-ready 
hyperparameter tuning system that is agnostic of the underlying
machine learning framework.
Though there are multiple hyperparameter tuning systems available, 
this is the first one that 
caters to the needs of both users and administrators of the system. 
We present the motivation and design of the system and contrast it with existing
hyperparameter tuning systems,
especially in terms of multi-tenancy, scalability, fault-tolerance, 
and extensibility. 
It can be deployed on local machines, or hosted as a service 
in on-premise data centers, or in private/public clouds.
We demonstrate the advantage of our system using experimental results as well
as real-world, production use cases. Katib has active contributors 
from multiple companies and 
is open-sourced at 
\emph{https://github.com/kubeflow/katib} under the Apache 2.0 license.

\end{abstract}



\maketitle

\section{Introduction}
\label{introduction}
In machine learning, a hyperparameter is a parameter whose value must be fixed before the actual training process.
Consequently, hyperparameters (e.g., number of clusters in k-means clustering, 
learning rate, batch size, and number of hidden nodes in neural networks)
cannot be learnt during the training process,
unlike the value of model parameters (e.g., weights of the edges in the neural network). Hyperparameters can impact both the quality of the model generated by 
the training process as well as the time and memory requirements of the algorithm~\cite{Goodfellow-et-al-2016}.
Thus, hyperparameters have to be tuned to get the optimal setting for a given problem. 
This tuning can be done either manually or automatically. Manual tuning might suffice if the
problem setup is not expected to change, at least not frequently, and if the number of
hyperparameters is not large.
Since these assumptions are frequently violated for realistic problems, 
an automatic hyperparameter 
tuning mechanism is required to make the overall machine learning approach practical.

There are several hyperparameter tuning systems that already exist. Notable among them are
Optuna~\cite{optuna:kdd19}, Ray Tune~\cite{tune:CoRR18}, Vizier~\cite{golovin:vizier-kdd2017}, 
HyperOpt~\cite{mypaper:hyperopt}, and NNI~\cite{mypaper:nni}. Even though these frameworks exhibit certain similarities (e.g., almost all frameworks support running parallel trials and customizable search algorithms),
there are important advantages that Katib has over these systems. Namely, Katib is the only open-source framework that can realistically be run as a hosted service in a production environment. 

There are several reasons for this distinction:
\begin{enumerate}
    \item \emph{Multi-tenancy}: The only other framework that supports multi-tenancy is Vizier, which is closed-source. Other frameworks support only single-tenant usage, which makes cross-team collaboration more difficult.
    \item \emph{Distributed Training}: Frameworks like Optuna and HyperOpt lack support for distributed training (either
parameter servers~\cite{nips2014:parameter_server} or 
collective communications such as RingAllReduce~\cite{horovod}). Both distributed patterns are supported in Katib.
    \item \emph{Cloud Native}: Katib is a Kubernetes-native~\cite{k8s} framework, thus making it a natural fit for cloud-native deployments. Other frameworks like Ray Tune and NNI support Kubernetes but require additional effort to configure.
    \item \emph{Extensibility}. Katib is an open-source framework with pluggable interfaces for search algorithms and data storage. Some frameworks (like HyperOpt) do not support customizing search algorithms, and most do not support customizing the underlying metric storage.
\end{enumerate}




Finally, Katib is one of only two frameworks that currently support neural
architecture search~\cite{katib:opml19} (the other one being NNI). Neural architecture
search (NAS) is an advanced automated machine learning technique that constructs 
entire neutral networks using various search strategies. A more detailed comparison of 
various frameworks is shown in Table~\ref{tab:frameworks}.


In addition to these differences,
Katib is the first system that was designed from the ground up
with a focus on multiple persona of users. Thus, the system does not simply cater to 
a data scientist persona (called a \emph{user} here), as do all the prior frameworks, 
but specifically targets the operations
persona (called an \emph{admin} here) as well.
Consequently, among all the competing systems in open-source,
Katib has the most support and contribution from more than 20 companies,
many of whom are 
already using it in production. Katib is open-sourced at 
\url{https://github.com/kubeflow/katib} under the Apache 2.0 license.

Our contributions in this paper are as follows:
\begin{enumerate}
    \item Katib is the first hyperparameter
    tuning system that is completely open-source and supports multi-tenancy,
    scalability, and extensibility, thus making it 
    the only candidate for a hosted hyperparameter tuning service.
    \item Katib is the only hyperparameter tuning system that was designed from the ground up
    with a focus on the usability for different persona of users,
    thus catering to both the data scientist as well as the administrator of the system.
\end{enumerate}

The rest of this paper is arranged as follows. Section~\ref{motivation} goes into how accounting 
for multiple personas leads to a unique set of requirements. Section~\ref{design} goes into
the detailed design and system workflow, followed by a list of supported features in
Section~\ref{features}. We present evaluation in Section~\ref{evaluation} and finally
conclude in Section~\ref{conclusion}.

\section{Motivation}
\label{motivation}
In order for a system to be deployed in an production setup, it needs to cater to multiple
personas of users. From the point of view of a hyperparameter tuning system, 
we can broadly classify into \emph{two} personas, \emph{user} and \emph{admin},
both having a different set of expectations from the underlying system.

The {\user}, often called the data scientist or machine learning engineer,
focuses on building, testing, and maintaining production ready machine learning models. 
The user is interested in developing the best performing machine learning models
using all available frameworks and tools in the market. The primary requirements for the
user are as follows:
\begin{enumerate}[label=(U.\arabic*)]
\item In the preliminary phase, the user wants to do HP tuning in a limited 
resource environment, e.g., laptop.
\item Once promising results are obtained from the initial experiments, 
the user wants to try a similar experiment in a much larger compute environment,
possibly with support for accelerators like GPUs and TPUs.
\item The user wants to compare and visualize the results after HP tuning.
\item The user wants to track, version, and share the experiment details and results.
\end{enumerate}

The {\admin}, also known as the operations engineer, has a vastly different set of 
requirements. The admin mainly
focuses on maintaining hardware and software platforms
which can be on premise or in the cloud. 
This persona
is responsible for managing the underlying infrastructure, maintaining its health,
and ensuring that the software on the infrastructure is highly available and up to date.
The admin needs the ability to:
\begin{enumerate}[label=(A.\arabic*)]
\item Perform resource efficient deployments.
\item Support multiple users in the same cluster with
dynamic resource allocation.
\item Share the cluster with non hyperparameter tuning jobs.
\item Perform capacity planning based on the user workloads with ease of cluster management.
\item Upgrade live system without affecting other users or running workloads.
\item Identify issues in the system through logging and monitoring.
\end{enumerate}

The design of Katib, from the very beginning, was heavily influenced by
these two different set of requirements of the two personas. To the best of our
knowledge, existing hyperparameter tuning systems
have primarily catered to the \emph{user} persona,
while largely neglecting the \emph{admin}.

\section{Design}
\label{design}

Some of the requirements of the {\admin} persona, 
specified in the previous section, are not unique
to a hyperparameter tuning system but are indeed the requirements for operating any
scalable, cloud-native service. Increasingly, such large scale systems 
are built using containers, coupled with an orchestration system for deploying, 
scaling, and managing containers.
The de-facto standard container orchestration system used today is
Kubernetes~\cite{k8s}, which is an open-source software used for 
``production-grade container orchestration''. 
Consequently, from the very beginning, the design of Katib was 
tightly coupled with Kubernetes, thus reusing a large set of existing tools and ensuring
compatibility to a rich cloud-native ecosystem. Next, we present some key definitions 
that are critical for explaining the design of the Katib system.

\subsection{Definitions}
\label{definitions}

The following are standard Kubernetes concepts and are included here 
for the sake of completeness:
\begin{itemize}
\item Node -- A single machine which can be a physical or virtual entity defined with a
set of resources like  memory, CPU, etc.

\item Cluster -- A collection of nodes which represents the distributed deployment setup. 
It can be deployed either on a local machine, on-premise data centers,
or private/public clouds.   
   
\item Resource -- A Kubernetes persistent object. The configuration for a resource 
is described in yaml format that 
contains two nested fields -- ``Spec'' and ``Status''. ``Spec'' refers to the
desired state of the object while ``Status'' refers to the current state. 
\end{itemize}

In addition to the above, some fundamental Katib constructs are as follows:
\subsubsection{Experiment}
\label{experiment}
An external (user-facing) resource referring to one complete user run 
or an optimization loop for a 
specific machine learning model. The experiment specifies user training task definition,
the objective to be optimized, the
parameter search space, and a search algorithm to be used.
 
\begin{table}[t]
\begin{tabular}{p{0.45\textwidth}}
    {
\begin{lstlisting}[language=yaml, numbers=left, linerange={1-53}, caption=Experiment Specification, label=spec:experiment]
objective:
    type: maximize
    goal: 0.99
    objectiveMetricName: Validation-accuracy
    additionalMetricNames:
    - accuracy
parameters:
    - name: --lr
      parameterType: double
      feasibleSpace:
        min: "0.01"
        max: "0.03"
    - name: --num-layers
      parameterType: int
      feasibleSpace:
        min: "2"
        max: "5"
    - name: --optimizer
      parameterType: categorical
      feasibleSpace:
        list:
        - sgd
        - adam
algorithm:
    algorithmName: bayesianoptimization
    algorithmSettings:
      - name: "random_state"
        value: "10"
parallelTrialCount: 3
maxTrialCount: 12
maxFailedTrialCount: 3

trialTemplate: |-
    apiVersion: batch/v1
    kind: Job
    metadata:
      name: {{.Trial}}
      namespace: {{.NameSpace}}
    spec:
      template:
        spec:
          containers:
          - name: {{.Trial}}
            image:  katib-mnist-example
            command:
            - "python"
            - "/classification/train_mnist.py"
            - "--batch-size=64"
            {{- with .HyperParameters}}
            {{- range .}}
            - "{{.Name}}={{.Value}}"
            {{- end}}
            {{- end}}
\end{lstlisting}
    }
\end{tabular}
\end{table}
A sample experiment specification is given in Listing~\ref{spec:experiment}.
The first section describes the `objective' to be optimized. In the example,
the target is to maximize the objective metric, \emph{Validation-accuracy} to reach a
goal value of 0.99.
An additional metric, \emph{accuracy} is also tracked in the experiment. 
Though multi variable optimization
is not supported at this point, multiple metrics can be tracked simultaneously, thus giving 
a better understanding of the tuning progress. 
The progress of these metrics over time can be viewed in the Katib dashboard
(shown later in Section~\ref{evaluation}). 
The second section describes the `parameters' to
be optimized with their type and search space. 
Supported types are \emph{integer, double, discrete,} and \emph{categorical}. 
The example specifies three parameters --- a double type \emph{lr}, 
an integer type \emph{num-layers}, 
and a categorical type \emph{optimizer}.
The third section describes the search `algorithm' to be used 
for searching hyperparameter values 
and its corresponding settings. In the example spec, the search algorithm used is 
Bayesian optimization 
with the setting \emph{random\_state} set to 10.
The fourth section describes the experiment-wide settings that control the whole run. 
`ParallelTrialCount' specifies the number of trials to be executed in parallel; 
`MaxTrialCount' specifies the experiment budget or maximum number of 
completed trials for an 
experiment to be marked successful. `MaxFailedTrialCount    '
specifies the error budget or maximum number of failed 
trials before an experiment is marked as failed.

The fifth section describes the `trialTemplate' which
defines the user training task to be optimized and follows the Go Template format. The example spec specifies an MNIST example container image 
with hyperparameters passed as command line arguments.

\subsubsection{Suggestion}
A suggestion is an internal resource (not exposed to the user),
referring to one proposed solution to the optimization problem or 
a set of generated hyperparameter values.

\subsubsection{Trial}
A \emph{Trial} is an internal
resource referring to one iteration of the optimization loop or an instance 
of the training job with generated suggestion values.

\subsubsection{TrialJob}
A \emph{TrialJob} is the training instance provided by the user.
A \emph{TrialJob} can be a non-distributed training job with a single worker, or a
distributed job consisting of several workers. Katib is designed to work with
Kubeflow~\cite{kubeflow} -- an open-source machine learning toolkit for Kubernetes --
and natively supports TensorFlow~\cite{tensorflow},
PyTorch~\cite{pytorch}, and XGBoost~\cite{chen2016xgboost} distributed training jobs.

\begin{lstlisting}[language=yaml,
                 linerange={14-32},
                 numbers=left,
                 caption=Trial Specification,
                 label=spec:trial]
runSpec: |-
  apiVersion: batch/v1
  kind: Job
  metadata:
    name: bayesian-run-1
    namespace: kubeflow
  spec:
    template:
      spec:
        containers:
        - name: bayesian-run-1
          image: katib-mnist-example
          command:
          - "python"
          - "/classification/train_mnist.py"
          - "--batch-size=64"
          - "--lr=0.013884981186857928"
          - "--num-layers=3"
          - "--optimizer=adam"
\end{lstlisting}
\subsubsection{Controller}
The controller is a non-terminating process that watches the state of a
resource and makes required changes
attempting to move the current state (Status) of a resource closer towards
the desired state (defined in Spec).
For example, the Experiment controller provides
life cycle management of the Experiment resource. 
Similarly, Trial and Suggestion controllers provides
life cycle management of the Trial and Suggestion resources respectively.

\subsection{System Workflow}
\label{workflow}

\begin{figure}[!tb]
\centering
\includegraphics[width=0.45\textwidth]{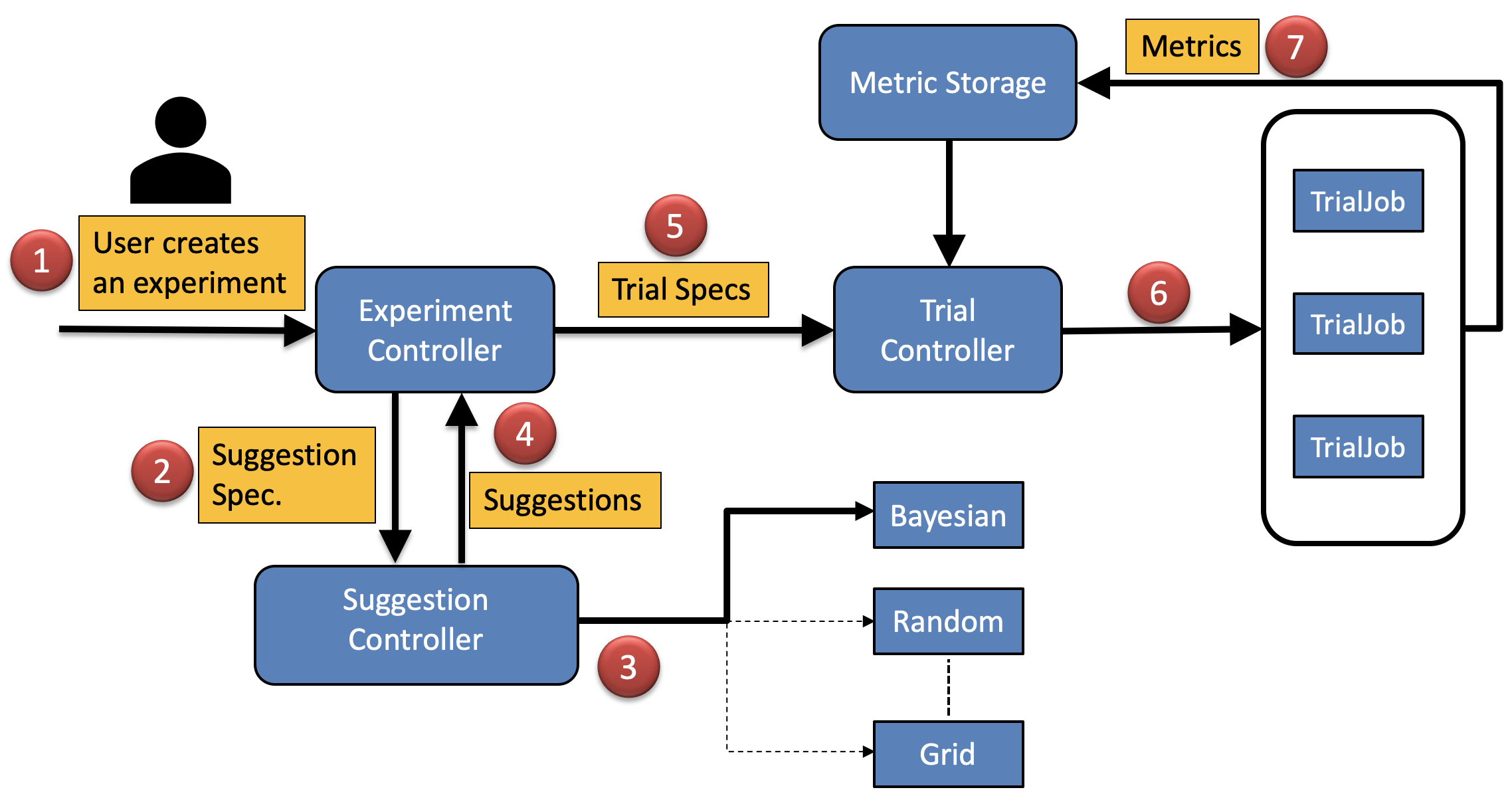}
\caption{Katib System Workflow.
}
\label{fig:arch}
\end{figure}

In this section, we present the typical workflow that a {\user} 
would follow to interact with the Katib system. The schematic of the workflow
is presented in Figure~\ref{fig:arch} and the major steps are as follows:

\begin{enumerate}[leftmargin=12pt]  
\item The {\user} creates an Experiment Spec in Yaml
and submits it to the Katib system using client tools. 
Since OpenAPI~\cite{openapi} specification is also supported, the {\user}
can alternatively generate
clients based on their choice of language and construct the config.
For better clarity
of the workflow, we will use experiment spec defined in Listing~\ref{spec:experiment}
as an example. 

\begin{lstlisting}[language=yaml,
                 numbers=left,
                 linerange={1-23},
                 %caption=Suggestion Specification,
                 caption=Suggestion Resource,
                 label=spec:suggestion]
spec:
  algorithmName: bayesianoptimization
  requests: 2
status:
    suggestionCount: 2
    suggestions:
    - name: bayesianoptimization-run-1
      parameterAssignments:
      - name: --lr
        value: "0.013884981186857928"
      - name: --num-layers
        value: "3"
      - name: --optimizer
        value: adam
    - name: bayesianoptimization-run-2
      parameterAssignments:
      - name: --lr
        value: "0.024941501303260026"
      - name: --num-layers
        value: "4"
      - name: --optimizer
        value: sgd
\end{lstlisting}  
 
\item The experiment controller reads the Experiment Spec and 
creates the corresponding Suggestion spec as given in the
Listing~\ref{spec:suggestion}. Suggestion Spec specifies the search algorithm
and the number of requested suggestions. The number of requested suggestions is equal to the maximum number of trials to be executed in parallel. This is determined by the parameter `ParallelTrialCount' in the Experiment Spec. Hence, 2 suggestions are requested using the Bayesian optimization algorithm.

\item Based on the search algorithm parameter defined in the Experiment Spec,
an algorithm service is deployed per experiment.

\item The Suggestion Controller reads
Suggestion Spec and updates the Suggestion Status with requested number
of suggestions from the algorithm service. As seen in Listing~\ref{spec:suggestion},
Suggestion status has two sets of parameter assignments which is returned from the
deployed Bayesian optimization algorithm service.

\item The Experiment Controller reads the Suggestion Status and spawns 
multiple Trials with
each Trial Spec corresponding to one generated suggestion.
The trial template of the experiment specification shown in Listing~\ref{spec:experiment} gets converted to a run time
trial specification shown in Listing~\ref{spec:trial} for each generated suggestion. 
The `runSpec' in the Trial Spec is obtained by executing the
Trial template 
and replacing the  template variables \emph{name, namespace} and \emph{HyperParameters}
with parameter assignments from the generated suggestion based
on Listing~\ref{spec:suggestion}.

\item The Trial Controller reads the Trial Spec of each Trial and
creates corresponding TrialJobs.
The hyperparameters are passed to the training code through command line arguments. 

\item The Trial Controller continuously watches the status of all
spawned TrialJobs and updates the Trial Status. 
When underlying jobs are completed, the Trial's status is marked as completed.
Once completed, the TrialJob metrics are also reported to the underlying 
\emph{metric storage} and the best objective metric value is recorded in the 
Trial Status. 

\item The Experiment Controller reads the status of all Trials and verifies
if the experiment `objective' is met. 
If the experiment objective is met, the experiment is marked as complete. 

If the experiment objective is not met, steps 2-8 are repeated till the
configured experiment budget is reached.  

\end{enumerate}

\begin{table*}[!t]
    \resizebox{\textwidth}{!}{%
\begin{tabular}{||p{1.35in}||p{0.60in}|p{0.80in}|p{0.70in}|p{0.75in}|p{1.05in}||p{0.55in}||}
\hline
\textbf{Framework} & \textbf{Optuna}~\cite{optuna:kdd19} & \textbf{Ray Tune}~\cite{tune:CoRR18} & \textbf{Vizier}~\cite{golovin:vizier-kdd2017} & \textbf{HyperOpt}~\cite{mypaper:hyperopt} & \textbf{NNI}~\cite{mypaper:nni} & \textbf{Katib}~\cite{katib} \\
    \hline
    \hline

\textbf{Open Sourced} & MIT & Apache 2.0 & No & Custom & MIT & Apache 2.0 \\ \hline

\multicolumn{7}{||l||}{\textbf{Generic}}\\ \hline

\hspace{3mm}Language Agnostic & Python & Python  & Any & Python & Python & Any\\  \hline
\textbf{Cloud Native} & Partial & Partial & No & No & Partial & Yes \\ \hline

\textbf{Platform} & None & Ray, Kubernetes & Google Borg & None & Kubernetes, PAI~\cite{openpai}& Kubernetes \\ \hline
\textbf{Multi-Tenancy} & No & No & Yes & No & No & Yes \\ \hline
\multicolumn{7}{||l||}{\textbf{Scalability}}\\ \hline
\hspace{3mm}Autoscalability & No & Yes & Yes & Partial & Partial & Yes  \\ \hline
\hspace{3mm}Distributed Execution & No & Yes & Yes & No & Yes & Yes \\ \hline
\multicolumn{7}{||l||}{\textbf{Extensibility}}\\ \hline
\hspace{3mm}User Code Invasiveness & High & High & Low & High & Low & Low \\ \hline
\hspace{3mm}Metric Storage & Partial & No & Partial & No & Partial & Yes \\ \hline
\hspace{3mm}Metrics Collection & Push & Push & Push & Push & Push & Pull/Push                                            \\ \hline
\hspace{3mm}Search Algorithm & Yes & Yes & Yes & Partial & Yes & Yes \\ \hline
\multicolumn{7}{||l||}{\textbf{Fault tolerance}}\\ \hline
\hspace{3mm}Trial Failure & No & Yes & Yes & No & No & Yes \\ \hline
\hspace{3mm}Error Budgets & No & No & No & No & No & Yes \\ \hline
\textbf{NAS} & No & No & No & No & Yes & Yes \\ \hline

\textbf{Gang Scheduling~\cite{gang-scheduling}} & No & No & No & No & No & Yes \\ \hline

\hline
\end{tabular}
}
\caption{Comparison of different hyperparameter tuning frameworks. Vizier has a close-source implementation
thus making it difficult to compare against.}
\label{tab:frameworks}
\end{table*}

In addition to this workflow, the {\admin} carries out a host of activities,
as specified in Section~\ref{motivation} (requirements A.1 -- A.6).
These tend to have separate workflows that are usually not specific to the
Katib system but are well known workflows
more tied to the administration of the underlying Kubernetes
infrastructure.

\section{Features}
\label{features}

Having presented the design of Katib, in this section,
we elaborate on the main features from Table~\ref{tab:frameworks}; the table
also contrasts Katib with comparable systems. These set of features span the
requirements of both the {\user} and the {\admin} personas, as described in
Section~\ref{motivation}.

\subsection{Generic}
Katib is \emph{framework agnostic} to machine learning frameworks. It can tune hyperparameters of applications written in \emph{any}
language of the users' choice and natively supports many machine 
learning frameworks, such as TensorFlow, PyTorch, MPI, and XGBoost.


\subsection{Multi Tenancy}
Katib provides multi tenancy using namespaces and access control rules. 
A namespace is defined to be a logical separation of cluster resources.
Each user is assigned a unique namespace.
The user can create, modify, view, and delete experiments in the assigned namespace.
In the default case,
access rules are automatically set for each user that prevents
unauthorized access to other namespaces.
However, the {\admin} can grant extra permission to users that permits
shared access to multiple namespaces.
This is particularly useful for users who want to collaborate across teams.  

\begin{figure}[!t]
\centering
\includegraphics[width=0.45\textwidth]{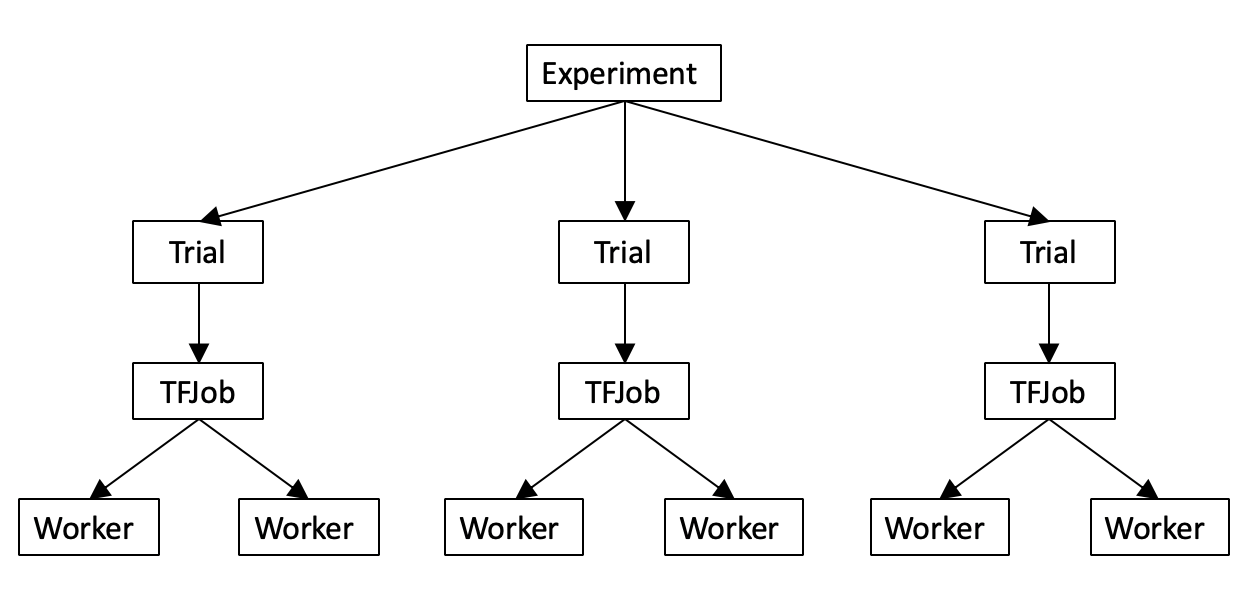}
\caption{Distributed TensorFlow job with 2 workers and ParallelTrialCount=3
runs 3 parallel trials with 2 workers each.}
\label{fig:distributed}
\end{figure}

Namespaces also provide the ability to set limits on resources like memory, CPU, and GPU.
For better capacity planning of the deployment cluster,
the admin defines in advance the limit for each user to ensure that cluster
resources are not oversubscribed.
Resource limits can be also set at individual \emph{experiment} level, thus
restricting the resource usage per experiment.

\subsection{Scalability}

Katib allows \emph{distributed execution} at multiple levels to provide cloud scale.
The experiment execution can be distributed at Trial as
well as at TrialJob level.
`ParallelTrialCount' parameter determines the amount of parallelism at the trial level.
For example, if `ParallelTrialCount' is set to 10, 10 trials can run in parallel,
subject to resource limits.
TrialJob execution can be non-distributed or distributed based on the framework
that the trial uses. For example,
TensorFlow, PyTorch, XGBoost, and MPI frameworks support distributed training jobs.
Trial and TrialJob parallelism can be simultaneously tuned for maximum
resource utilization in their deployment environment;
an example is shown in Figure~\ref{fig:distributed}.

Katib also supports \emph{Auto Scalability}, which allows cluster size to be 
automatically adjusted so that there is
no under- or over-utilization of resources.
In the autoscaler configuration, the minimum and maximum number of nodes can be
set. The cluster is
automatically scaled up when there are jobs
that cannot be scheduled in the cluster due to insufficient
resources. Similarly, the cluster is automatically scaled down when
there are nodes left underutilized for a configurable period of time.
This helps in controlling the total cost without exceeding the
target budget for the experiment runs.
Since Katib is \emph{horizontally scalable}, new nodes can be added or removed from the 
cluster during run time.

\subsection{Extensibility}
\label{extensibility}
\begin{figure}[!t]
\centering
\includegraphics[width=0.45\textwidth]{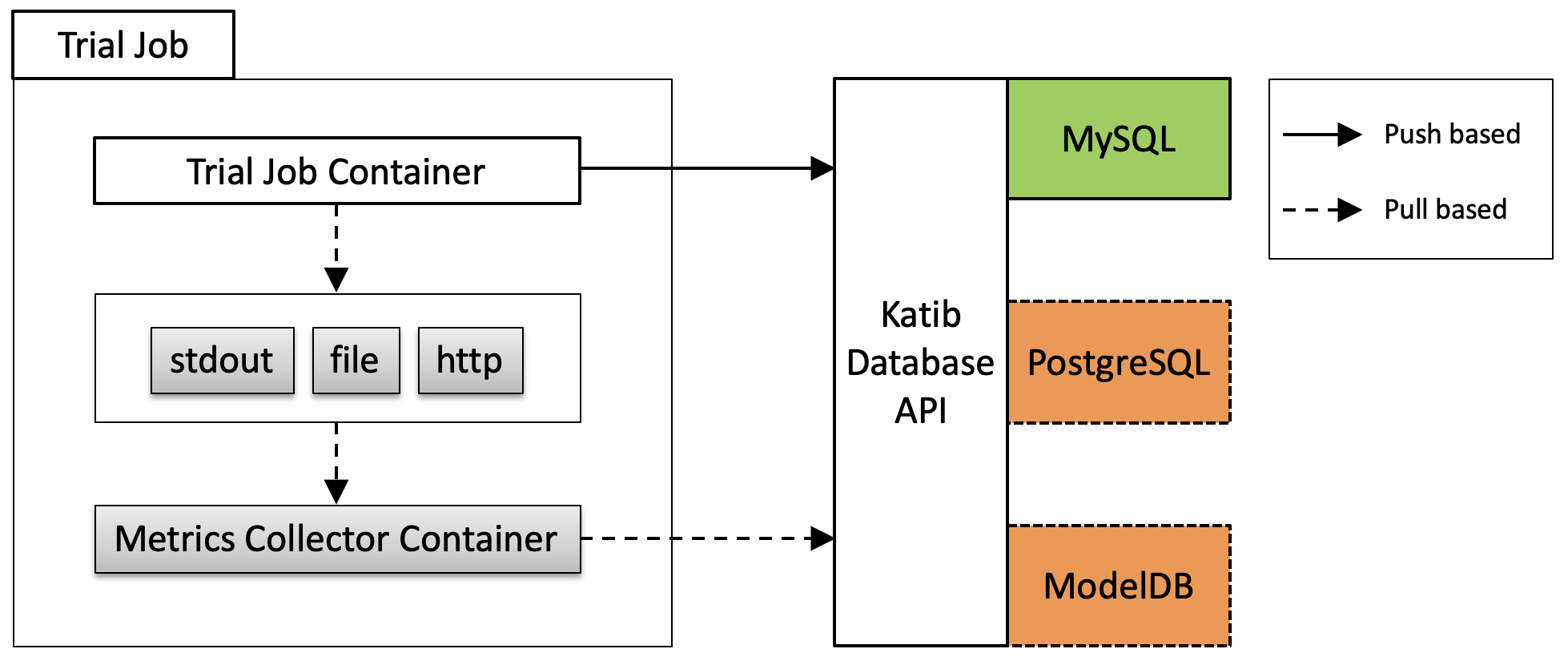}
\caption{Push and pull based metric collection in Katib.}
\label{fig:metric_collection}
\end{figure}

Katib exposes a pluggable database interface for different types of \emph{metric
storage}, as shown in Figure~\ref{fig:metric_collection}.
Any database can be supported in Katib by implementing the following functions
of the Katib Database API:
\begin{itemize}
    \item \emph{RegisterObservationLog()} -- Save trial metrics into database.
    \item \emph{GetObservationLog()} -- Retrieve trial metrics from database based on filters like start-timestamp and end-timestamp.
    \item \emph{DeleteObservationLog()} -- Delete trial metrics from database.    
\end{itemize}
    
The database can be either a local deployment or a remotely hosted service
like Amazon Relational Database Service (RDS). 
Currently, MySQL~\cite{mysql} is the default deployed database.
There are ongoing efforts to support PostgreSQL~\cite{postgresql}, 
ModelDB~\cite{modeldb}, Kubeflow Metadata DB~\cite{kubeflow},
and MLFlow~\cite{mlflow}.

As shown in Figure~\ref{fig:metric_collection},
Katib supports two kinds of metric collection --
\emph{push based} and \emph{pull based},
which can be configured using the
\emph{metricCollectorKind} parameter in the experiment specification.
In push based metric collection, the metrics are pushed directly from the training container
to the underlying \emph{metric storage} using database 
APIs described above.
The user training code has to be modified accordingly to enable metric tracking. 
In contrast, the pull based metric collection works with unmodified training code but
it doesn't provide synchronous control over the metrics written to the database.
In this approach, there is a 
sidecar container (a sidecar is a container that does not exist by itself but is
always paired with a \emph{main} container)
which pulls logs from the training container,
applies custom parsing and then pushes the metrics to the underlying \emph{metric storage}.

Katib also exposes pluggable algorithm interface to support new hyperparameter
\emph{suggestion algorithms}. This allows the user 
to plug in any custom algorithm that is suited to their environment needs. Any new \emph{suggestion algorithm} can be integrated by
implementing the \emph{GetSuggestions()} API, which generates a new suggestion of hyperparameters
to be evaluated.
Currently, the supported algorithms are Random~\cite{random:jmlr12}, Grid~\cite{random:jmlr12}, Bayesian Optimization~\cite{bayesopt:nips12},
Hyperband~\cite{hyperband:jmlr17}, and TPE~\cite{tpe:nips11}.

\subsection{Fault Tolerance}
Since Katib leverages Kubernetes, it is a highly available system without
any single point of failure. 
The latest state of any resource is recorded in its status field.
Upon node restart or failure, 
jobs are redeployed automatically, resuming them from the last recorded checkpoint. 

\subsection{Portability}
Katib system offers high portability,
allowing the {\user} to execute workloads across different environments with minimal effort.
Because Katib is designed to be cloud-native, its resource management and 
scheduling mechanisms are decoupled from the underlying infrastructure. 
Consequently, users can deploy and reproduce Katib experiments on a variety of environments,
such as on a laptop, in a private cluster, or on a public cloud. 
The user only needs to define the total budget and run-time requirements for each trial,
and the underlying system takes care of the actual resource allocation.

\subsection{Upgradeability}
Since Katib is designed to be cloud native, all individual components listed in Figure~\ref{fig:arch} are loosely coupled, allowing the live upgrade of components 
without any downtime. Since all components are versioned,
they can be independently upgraded 
without affecting other components. Existing experiments can be smoothly upgraded to a
newer version in a live cluster. Similarly, newer algorithms can be added
during runtime without affecting 
any running experiments. If existing algorithms are updated, the changes are reflected only from the
next experiment.

\subsection{NAS}
Katib system is designed to be a general automated machine learning platform supporting features such as Neural Architecture Search (NAS).  Since Katib follows an extensible architecture as shown in Figure~\ref{fig:arch},several classes of NAS algorithms can be easily supported. Currently, it supports NAS based on reinforcement learning strategy. Since it is already presented in a 
previous paper~\cite{katib:opml19},  we do not go into much detail.

\section{Evaluation}
\label{evaluation}

In this section, we compare Katib with some other existing hyperparameter tuning 
systems and demonstrate features
of Katib through some experiments. Then we introduce some 
real world applications in industries.


\subsection{Feature Evaluation}

\subsubsection{Portability}

\begin{figure*}[!t]
\centering
\subcaptionbox{Laptop: 15 trials on a Minikube cluster.\\ Optimal accuracy (97.7\%) is obtained with SGD optimizer,\\learning rate 0.212, 3 layers, and batch size 800\label{fig:port:a}}
    {\includegraphics[width=0.45\textwidth]{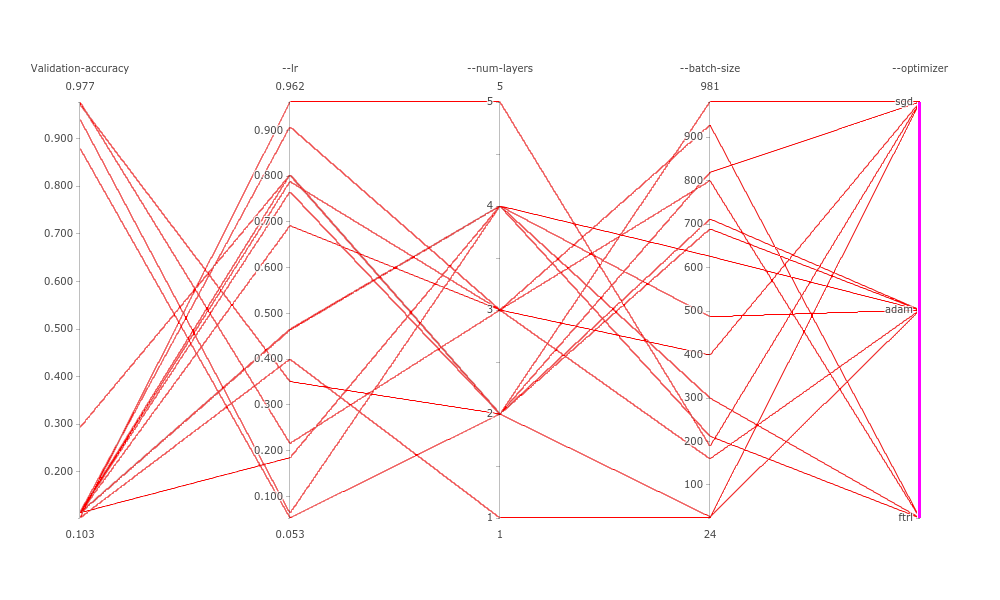}}
\subcaptionbox{Cloud: 50 trials on 16vCPU cores on GKE cluster.\\ Optimal accuracy (98.3\%) is obtained with SGD optimizer,\\
learning rate 0.287, 4 layers, and batch size 985.\label{fig:port:b}}
    {\includegraphics[width=0.45\textwidth]{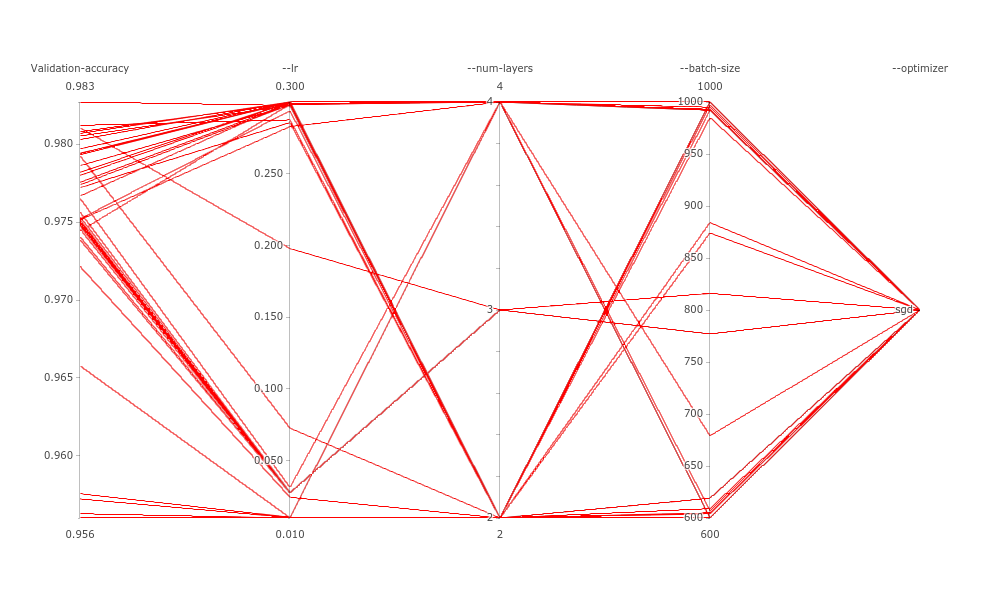}}
\caption{Portability in Katib\label{fig:portability}. Preliminary search is run
on a laptop followed by extensive tuning in the cloud.}
\end{figure*}

In this experiment, we demonstrate that Katib is a highly portable system
that can run on various platforms with minimal configuration changes. We simulate
what a typical user would do: run some preliminary trials locally on
a laptop, and then run a larger experiment with more trials on a cloud cluster.

We conducted the experiment with a simple MNIST model defined via Apache MXNet~\cite{mxnet2015}. The first part
of the experiment was run on a Minikube cluster (a single node Kubernetes cluster on a VM)
deployed on a laptop. We ran 15 trials with random search over the following hyperparameter ranges:
\begin{enumerate}
\item Learning rate: a float value ranging from 0 to 1.0
\item Batch size: an integer value ranging from 10 to 1000
\item Number of layers: an integer value ranging from 1 to 5
\item Optimizer: choice of SGD, Adam, and FTRL.
\end{enumerate}

The results are shown in Figure \ref{fig:port:a}. The leftmost axis ("validation-accuracy") shows the objective metric used to assess the trials, while the other axes plot the hyperparameter values used. Since we were using random search over fairly large ranges for hyperparameter values, the validation accuracy varied greatly as expected.
But we can see that the more promising results (validation-accuracy > 95\%) 
were obtained when using the SGD optimizer and learning rate at lower than 0.3.

In the second part of the experiment, the same experiment was ported to a GKE cluster
with 16 vCPU(virtual CPUs) cores. This time we made a few modifications: we changed
the search algorithm to Bayesian optimization and increased the number
of trials to 50. Also, based on the previous experiment, we narrowed
down the search space for the hyperparameters to:

\begin{enumerate}
    \item Learning rate: Between 0 and 0.3
    \item Batch size: Between 600 and 1000
    \item Number of layers: Between 2 and 4
    \item Optimizer: Only use SGD
\end{enumerate}

The results can be seen in Figure \ref{fig:port:b}. Almost all of the trials had resulting validation accuracy above 97\%, with the most optimal trial producing an accuracy of 98.3\%. 

Most hyperparameter tuning systems support the ability to run on multiple platforms, but in Katib exporting and reproducing experiments are very lightweight. This is because Katib is built to be Kubernetes-native, so the underlying infrastructure is completely abstracted away from the user persona. 
\emph{A Katib experiment can be exported as just a YAML file.}

\subsubsection{Multi Tenancy}
\begin{figure}[!t]
\centering
\includegraphics[width=0.4\textwidth]{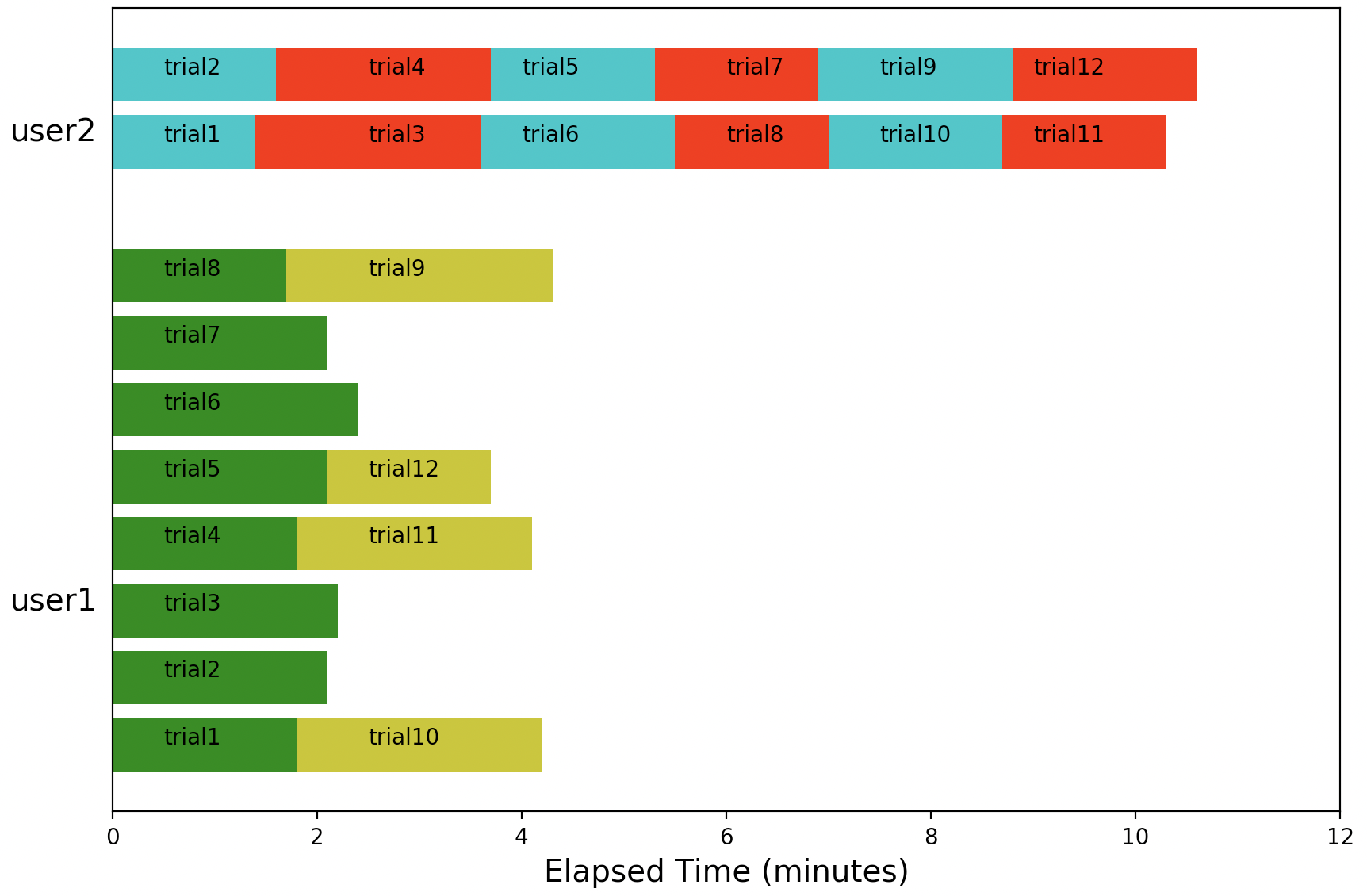}
\caption{Multi Tenancy using Katib.\label{fig:multi_tenancy}
}
\end{figure}

To demonstrate the multi tenancy feature,  we deployed Katib in a multi user environment.
Two users are configured in a 24 vCPU cluster with each having a separate namespace.
Resource quota is set for each namespace limiting the maximum CPU resources to 18
vCPUs for `user1' and 6 vCPUs for `user2'.
Each user is configured to run the same experiment config with MaxTrials set to 12, 
ParallelTrials set to 12, and each Trial requires 2vCPUs.  
The graph in Figure~\ref{fig:multi_tenancy} indicates that there are maximum 8 parallel trial executions for user1 and 2 parallel runs for user2
though ParallelTrials is configured to 12 for both users. 
Since aggregate resources required for parallel trials for each user cannot 
exceed the maximum resources allocated to the user in the assigned namespace, 
maximum executed trials in parallel is restricted to 8 and 2 respectively though the cluster can handle up
to 12 parallel trials.
Since suggestion algorithm is deployed for every experiment,
it takes 0.5 CPU by default from the available user resources, thus leaving only 1.5 vCPUs
which cannot be used to execute another trial since each trial
requires exactly 2 vCPUs.
To the best of our knowledge, Katib is the only open-source
hyperparameter tuning framework that natively supports multi-tenancy.

\subsubsection{Scalability}
\begin{figure}[!t]
\centering
\includegraphics[width=0.4\textwidth]{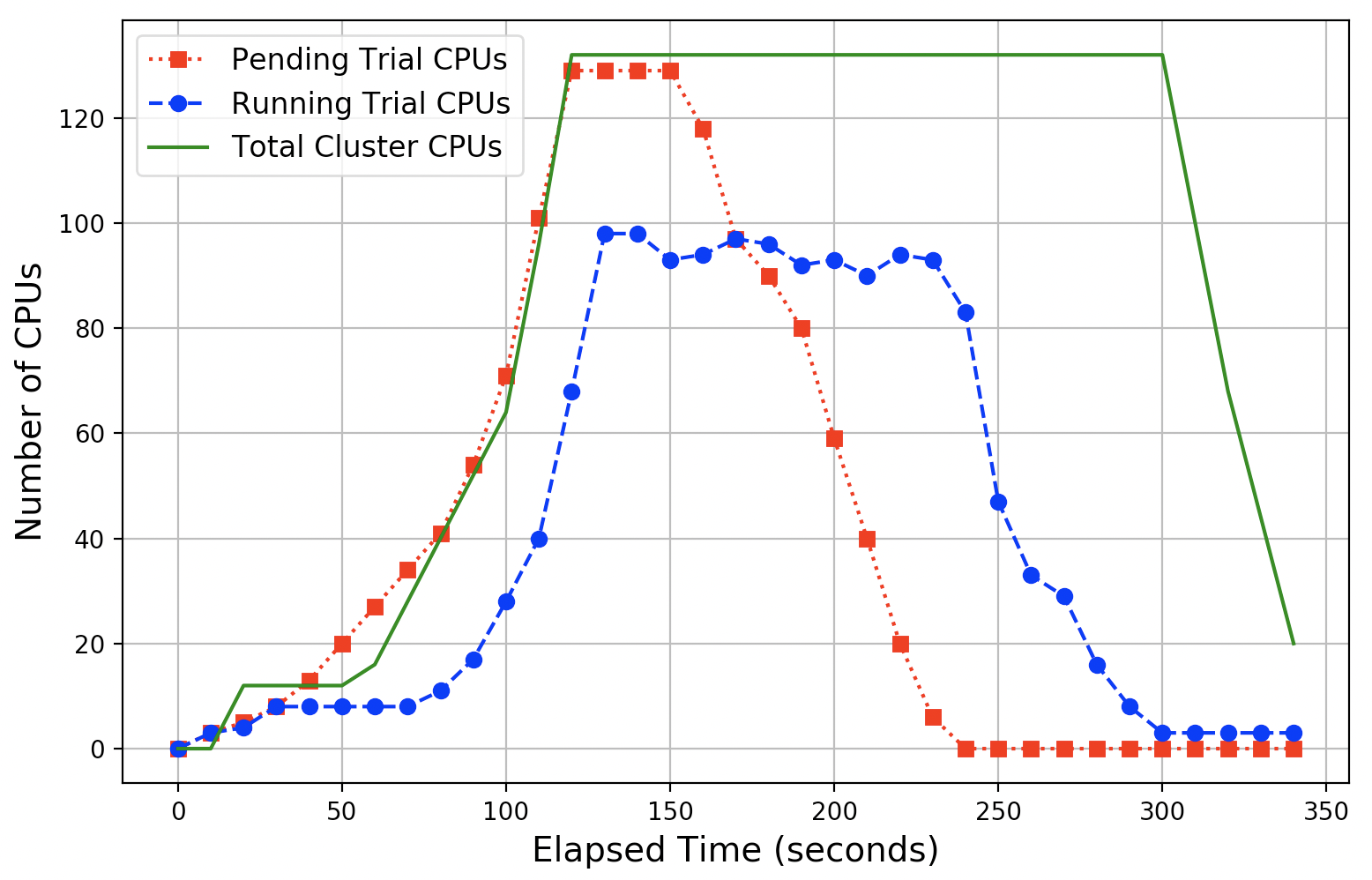}
\caption{Autoscaling in Katib.\label{fig:autoscale}}
\end{figure}

\begin{figure*}[!htbp]
\centering
\subcaptionbox{Trial failures vs Number of trials\label{fig:chaos:b}}
    {\includegraphics[width=0.33\textwidth]{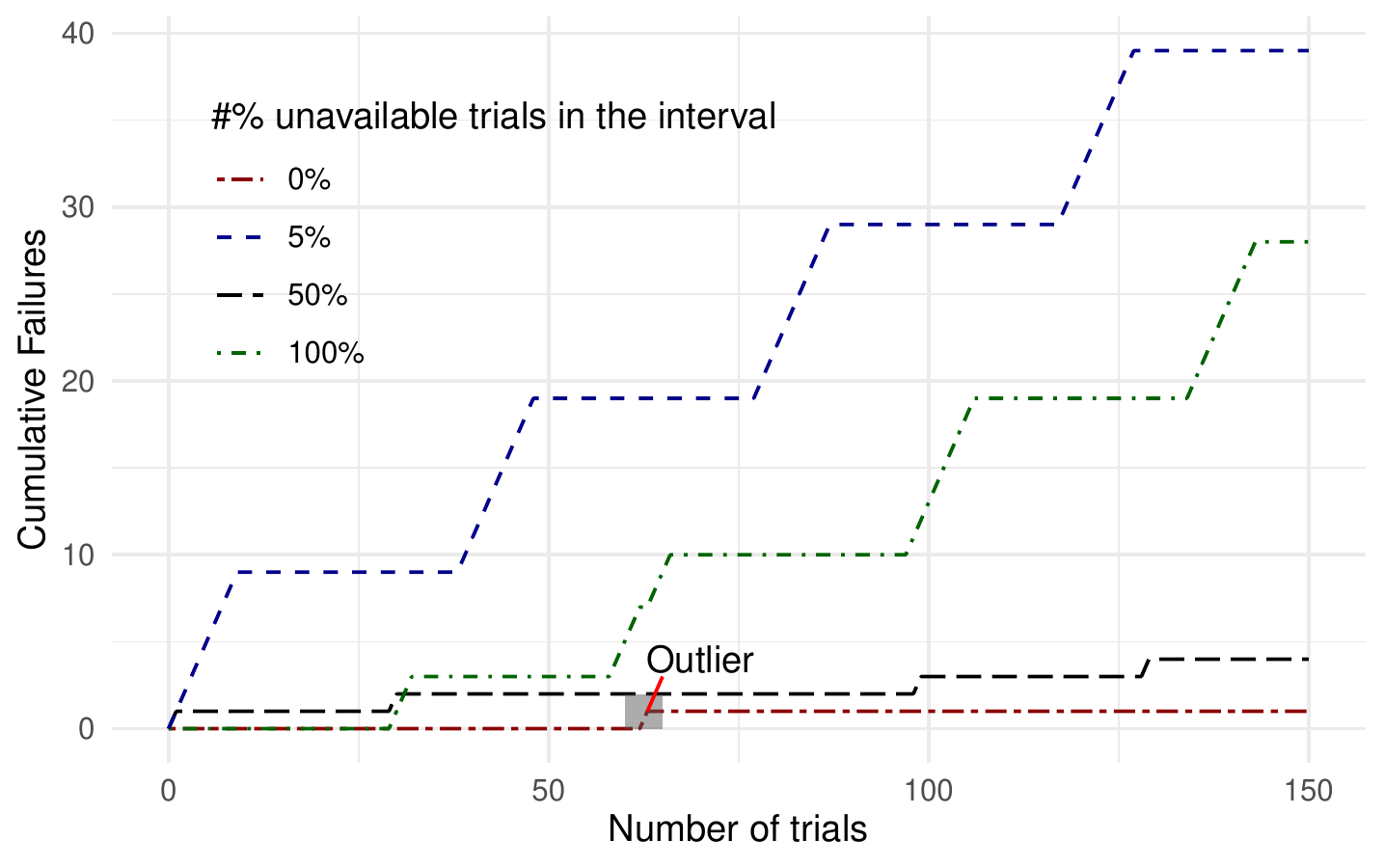}}
\subcaptionbox{Cross-entropy vs Number of trials\label{fig:chaos:a}}
    {\includegraphics[width=0.33\textwidth]{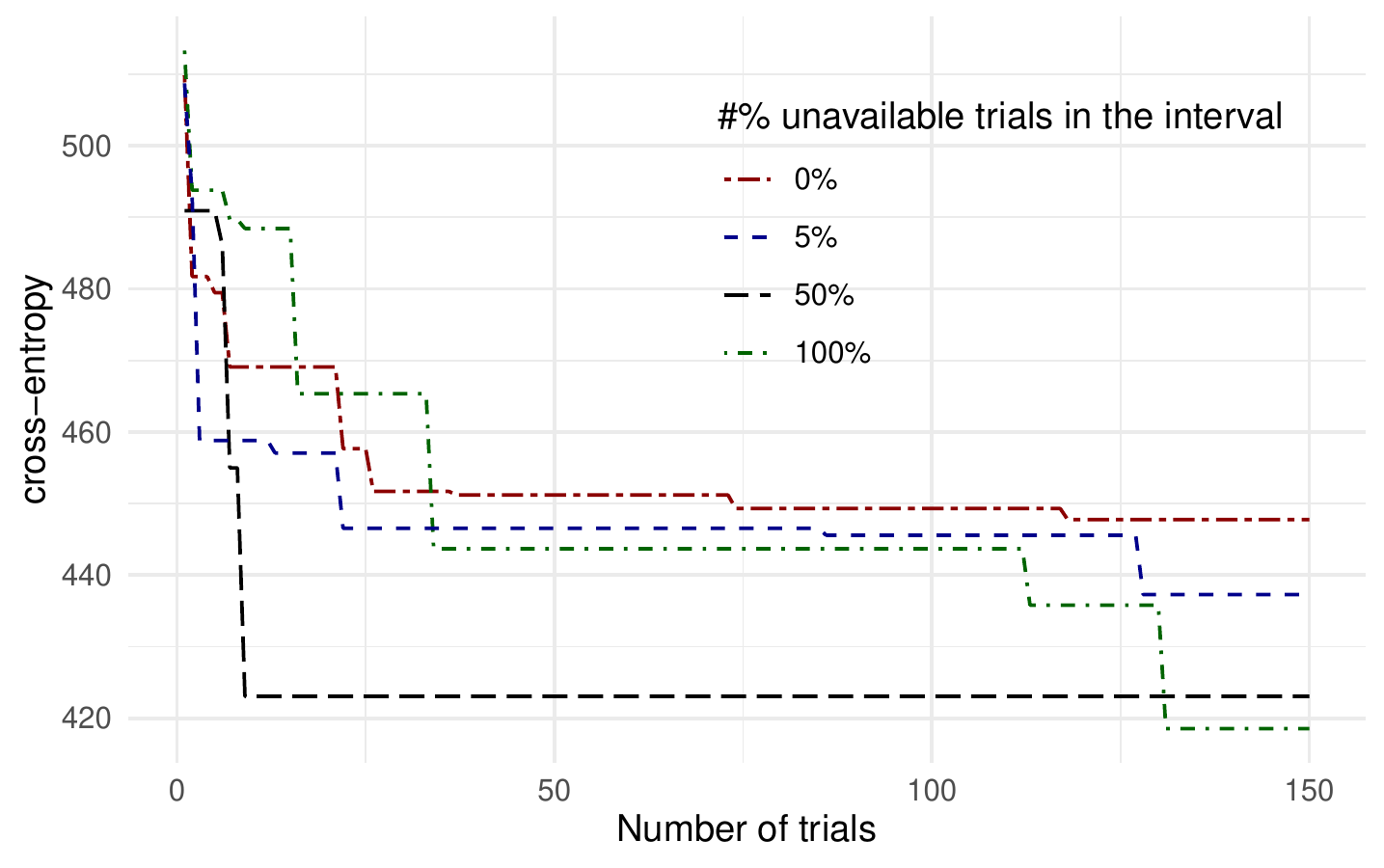}}
\subcaptionbox{Trial failures vs Number of trials\label{fig:kill}}
    {\includegraphics[width=0.33\textwidth]{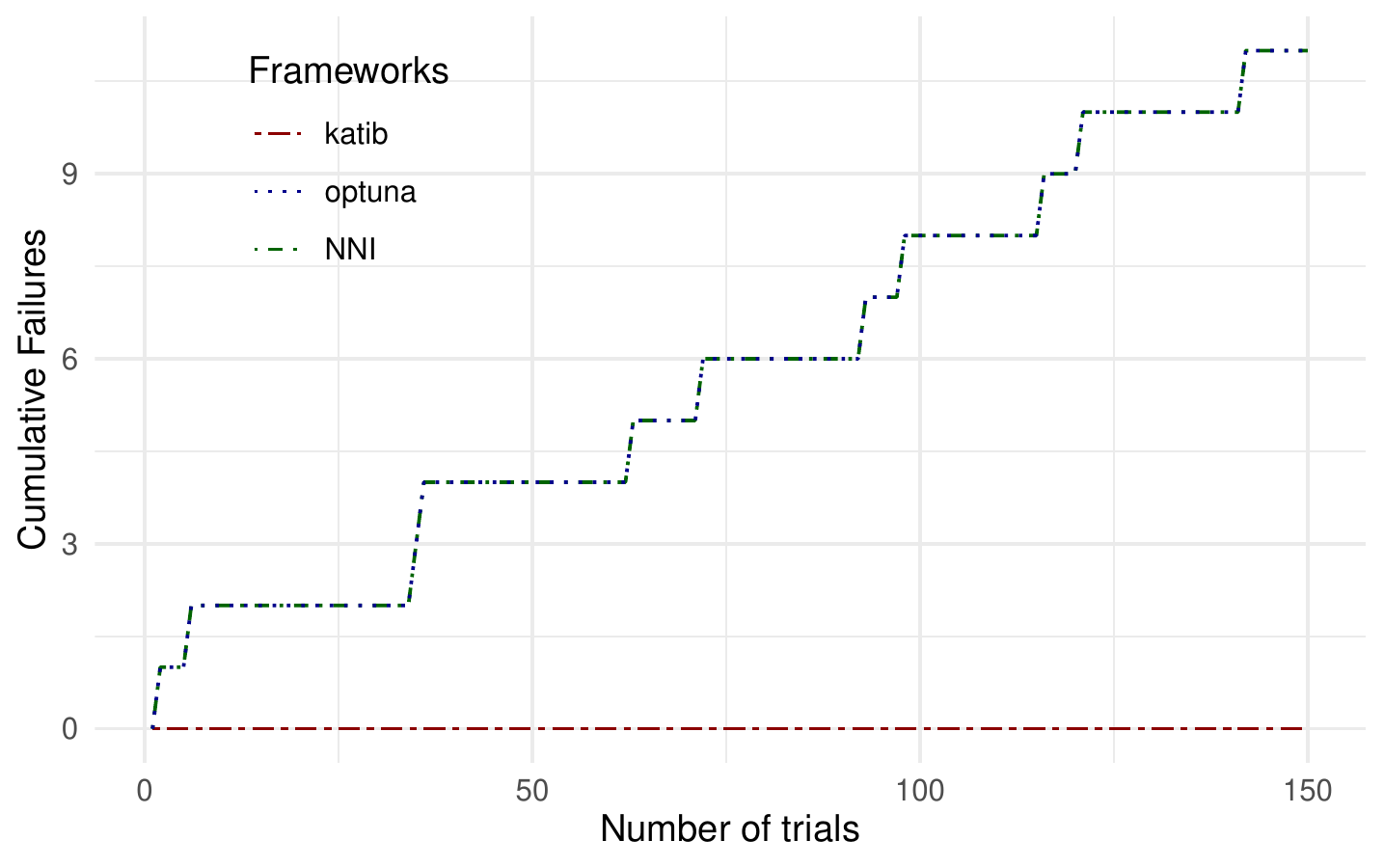}}
\caption{Fault Tolerance in Katib.}
\label{fig:chaos}
\end{figure*}

In order to evaluate scalability, Katib is deployed in a autoscaler enabled cluster with 3
nodes of 4vCPU each. The autoscaler is configured with minimum and maximum nodes as 3 and 50
respectively.  The experiment is run with MaxTrials and ParallelTrials set to 250 with each Trial
using 2 vCPUs each. Figure~\ref{fig:autoscale}
shows how autoscaler  automatically resizes the number of nodes
in the cluster based on the workload.  The cluster autoscaler adds extra nodes automatically
when there are pending Trials due to lack of CPUs. Figure also shows that cluster autoscaler
removes nodes automatically after some grace termination period when nodes are underutilized. This
ensures that total resource cost is controlled while ensuring the availability of the user
workload.  As indicated in the figure, the cluster autoscaler ensures that the required number
of Trial CPUs at any instant is within the limits of cluster capacity.

\subsubsection{Fault Tolerance}

Failures occur for a variety of reasons and not all failures are the same.
To inject failures, we applied chaos engineering\cite{basiri2016chaos} on 
Katib, Optuna, and NNI with the help of Chaos Mesh~\cite{chaos-mesh}, 
which is a cloud native chaos engineering platform. 
We designed couple of experiments to show how users can manage failures
with Katib and how it compares with other frameworks in this respect.

The Katib experiment spec is configured to minimize the objective metric, cross-entropy for a distributed Tensorflow Job with 2 workers. MaxFailedTrialCount is set to 100, MaxTrials set to 150, and ParallelTrials set to 10 for both experiments.

In the first experiment, a fixed proportion of Katib trials are failed at a fixed interval of time by the chaos engineering platform. 
Specifically, the platform is configured to fail 0\%, 5\%, 50\%, 
and 100\% of trials every 20 minutes and objective metric values of cross-entropy is 
collected from succeeded trials. Failures are simulated by altering the trial
container image names in the experiment spec to an invalid value which causes the 
trials to fail (and these trials cannot be recovered since the image names are invalid).
Figure \ref{fig:chaos:b} shows cumulative failures of trials for different failed trial ratios.
There are about 40 total failed trials when we fail all, i.e 100\%,  
trials every 20 minutes, but the hyperparameter exploration is not affected by these failures as indicated in Figure \ref{fig:chaos:a}. Figure \ref{fig:chaos:a} shows that objective metric values improve over time for all failure rates.
This indicate that failures do not have a huge 
impact on the performance of the hyperparameter tuning experiments, which makes Katib fault tolerant. 
Figure \ref{fig:chaos:b} shows a failed trial even for 0\% trial failure case because of a killed kubernetes process when its memory usage exceeds the limit. 

In the second experiment, we kill 5\% of trials every 20 minutes instead of 
failing them.
This is simulated by terminating one of the workers of the Tensorflow job.
We run the same experiment for Katib, Optuna, and NNI. 
The comparison results are shown in Figure \ref{fig:kill}. In Katib, no trials are marked failed because distributed TensorFlow training jobs in Katib
supports restarting or resuming the training job if the exit code of any worker indicates 
temporary failure.
In contrast, NNI and Optuna have failed trials since the trials created by them
cannot be restarted. We cannot run the same experiment for Ray Tune due to
the unavailability of tool to apply chaos engineering on Ray.

\subsection{Real World Applications}
\label{realworld}
Katib has been adopted in Ant Financial, Caicloud, Cisco, and many other enterprises. 
In this section, 
we present the applications in Ant Financial and Caicloud as examples.

\subsubsection{Hyperparameter tuning system at Ant Financial}

At Ant Financial, we manage Kubernetes clusters with tens of thousands of nodes~\cite{ant-financial-k8s-case-study} and have deployed Katib along with other Kubeflow operators.
One popular combination is to use Katib in conjunction with MPI Operator~\cite{mpi-operator}. 
The MPI Operator leverages the network structure and collective communication algorithms 
so that users don't have to worry about the right ratio between number of workers and parameter
servers to obtain the best performance. When used with Katib, users can focus on finding
reasonable hyperparameter search space of their chosen model architecture without spending time
on tuning the hyperparameters and the downstream infrastructure for distributed training. 

The models produced have been widely deployed in production and battle-tested 
in many different real life scenarios. 
One notable use case is Dingsunbao~\cite{mypaper:dingsunbao} -- a video-based mobile app that allows 
drivers to provide detailed vehicle damage information to insurers and claim vehicle
insurance in real time. Car owners can capture video streams of their cars on Dingsunbao app by following the on-screen guidelines. The system then uploads those captured video streams, recognizes vehicle damage information on the cloud asynchronously, and finally presents the damaged components to users automatically, with recommendations on where and how to repair the vehicle 
and how much the car
owner can claim from insurers. This makes filing claims easier without expensive laboratory costs and increases the
transparency in what's likely to be covered. Experiments have shown that the average damage assessment accuracy is 29.1\% higher and the ratio of high quality shooting data on predefined criterion is also 20\% higher compared with traditional approaches.

\subsubsection{Hyperparameter tuning as a service at Caicloud}
As a company focused on cloud-native machine learning infrastructure, we 
at Caicloud provide hyperparameter tuning services for customers in 
Caicloud Clever~\cite{caicloud-clever},
an artificial intelligence cloud platform. 
We implement a trial kind and a new \emph{metricCollectorKind} 
to integrate the metrics to Caicloud Clever.

Users can create hyperparameter tuning jobs in Caicloud Clever platfom.
Necessary source code, datasets or pretrained models will be pulled before
the actual Katib experiment run. Once the Katib
experiment is finished, the best-performing model and hyperparameter details are
pushed to the internal model registry. The model can be served and the experiment 
can be reproduced with the saved hyperparameters.

We also integrate the tuning service to the model marketplace in our platform.
There are some classical deep learning models available in the model marketplace. 
Users can import their datasets and tune the classical models with Katib 
without the need to build the models from scratch. 
In addition, there is also some ongoing research on how to run
advanced neural architecture search algorithms
such as DARTS\cite{DBLP:journals/corr/abs-1806-09055}
with Katib in order to automate the 
process for our customers.

\section{Conclusion}
\label{conclusion}

In this paper, we presented the motivation and design of Katib, a scalable and cloud-native 
hyperparameter tuning system that caters to both the {\user} and {\admin} personas. We 
contrast Katib with existing hyperparameter tuning systems and evaluate it
along several aspects that are critical to production-ready systems such as portability,
multi-tenancy, autoscaling, and fault tolerance.
We also present case studies of large-scale, real world applications that are
using Katib in production. 
Katib is a lively open-source project under the Apache 2.0 license and
has contributors from more than 20 companies.

\bibliographystyle{acm}
\bibliography{katib-arxiv}

\newpage
\appendix
\section{Reproducibility}
\label{reproduce}
In this section we provide some details regarding the reproducibility of the
experiments presented in the evaluation section of this paper.

\subsection{Environment}
All experiments were conducted on Google Kubernetes Engine (GKE) version 1.14.
The local cluster experiment was run using Minikube with KVM driver on a Linux server.
 
\subsection{Installation}
Katib installation guide is provided at
\url{https://github.com/kubeflow/katib\#installation}. 
Once Katib is installed, the configuration files can be submitted to the cluster 
using the Kubernetes command line tool kubectl. 

\subsection{Experiments}
The configuration and test files used in the evaluation section are uploaded to the
GitHub repository at \url{https://github.com/katib-examples/evaluation}. 
The test and configuration files for each evaluation 
section are put into corresponding folders in the GitHub repository:

\begin{enumerate}
\item Portability: 
\url{https://github.com/katib-examples/evaluation/tree/master/portability}

\item Multi Tenancy: 
\url{https://github.com/katib-examples/evaluation/tree/master/multi-tenancy}

\item Scalability:  \url{https://github.com/katib-examples/evaluation/tree/master/scalability}

\item Failure Tolerance:  
\url{https://github.com/katib-examples/evaluation/tree/master/fault-tolerance}

\end{enumerate}

\end{document}